\documentclass[twocolumn,letterpaper,aps,prc,longbibliography,superscriptaddress,showpacs,nofootinbib,floatfix]{revtex4-1}

\usepackage{graphicx}	
\usepackage{multirow}
\usepackage{xspace}	

\newcommand{\auau}{\mbox{Au$+$Au}\xspace} 

\newcommand{\pbpb}{\mbox{Pb$+$Pb}\xspace} 
\newcommand{\pdau}{\mbox{$p(d)$$+$Au}\xspace}

\newcommand{\pp}{\mbox{$p$$+$$p$}\xspace}

\newcommand{\sqrtsnn}{\mbox{$\sqrt{s_{_{NN}}}$}}

\newcommand{\pt}{\mbox{${p_T}$}\xspace}

\newcommand{\dAu}{\mbox{$d$$+$Au}\xspace}

\newcommand{\pythia}{\mbox{\sc pythia}\xspace}

\begin{document}

\title{$\phi$ meson production in the forward/backward rapidity 
region in Cu$+$Au collisions at $\sqrt{s_{NN}}=$200 GeV}

\newcommand{\abilene}{Abilene Christian University, Abilene, Texas 79699, USA}
\newcommand{\augie}{Department of Physics, Augustana University, Sioux Falls, South Dakota 57197, USA}
\newcommand{\banaras}{Department of Physics, Banaras Hindu University, Varanasi 221005, India}
\newcommand{\barc}{Bhabha Atomic Research Centre, Bombay 400 085, India}
\newcommand{\baruch}{Baruch College, City University of New York, New York, New York, 10010 USA}
\newcommand{\bnlcoll}{Collider-Accelerator Department, Brookhaven National Laboratory, Upton, New York 11973-5000, USA}
\newcommand{\bnlphys}{Physics Department, Brookhaven National Laboratory, Upton, New York 11973-5000, USA}
\newcommand{\caucr}{University of California-Riverside, Riverside, California 92521, USA}
\newcommand{\charlesczech}{Charles University, Ovocn\'{y} trh 5, Praha 1, 116 36, Prague, Czech Republic}
\newcommand{\chonbuk}{Chonbuk National University, Jeonju, 561-756, Korea}
\newcommand{\ciae}{Science and Technology on Nuclear Data Laboratory, China Institute of Atomic Energy, Beijing 102413, P.~R.~China}
\newcommand{\cns}{Center for Nuclear Study, Graduate School of Science, University of Tokyo, 7-3-1 Hongo, Bunkyo, Tokyo 113-0033, Japan}
\newcommand{\colorado}{University of Colorado, Boulder, Colorado 80309, USA}
\newcommand{\columbia}{Columbia University, New York, New York 10027 and Nevis Laboratories, Irvington, New York 10533, USA}
\newcommand{\czechtech}{Czech Technical University, Zikova 4, 166 36 Prague 6, Czech Republic}
\newcommand{\dapnia}{Dapnia, CEA Saclay, F-91191, Gif-sur-Yvette, France}
\newcommand{\elte}{ELTE, E{\"o}tv{\"o}s Lor{\'a}nd University, H-1117 Budapest, P{\'a}zm{\'a}ny P.~s.~1/A, Hungary}
\newcommand{\ewha}{Ewha Womans University, Seoul 120-750, Korea}
\newcommand{\fsu}{Florida State University, Tallahassee, Florida 32306, USA}
\newcommand{\gsu}{Georgia State University, Atlanta, Georgia 30303, USA}
\newcommand{\hanyang}{Hanyang University, Seoul 133-792, Korea}
\newcommand{\hiroshima}{Hiroshima University, Kagamiyama, Higashi-Hiroshima 739-8526, Japan}
\newcommand{\howard}{Department of Physics and Astronomy, Howard University, Washington, DC 20059, USA}
\newcommand{\ihepprot}{IHEP Protvino, State Research Center of Russian Federation, Institute for High Energy Physics, Protvino, 142281, Russia}
\newcommand{\illuiuc}{University of Illinois at Urbana-Champaign, Urbana, Illinois 61801, USA}
\newcommand{\inrras}{Institute for Nuclear Research of the Russian Academy of Sciences, prospekt 60-letiya Oktyabrya 7a, Moscow 117312, Russia}
\newcommand{\instpasczech}{Institute of Physics, Academy of Sciences of the Czech Republic, Na Slovance 2, 182 21 Prague 8, Czech Republic}
\newcommand{\isu}{Iowa State University, Ames, Iowa 50011, USA}
\newcommand{\jaea}{Advanced Science Research Center, Japan Atomic Energy Agency, 2-4 Shirakata Shirane, Tokai-mura, Naka-gun, Ibaraki-ken 319-1195, Japan}
\newcommand{\jyvaskyla}{Helsinki Institute of Physics and University of Jyv{\"a}skyl{\"a}, P.O.Box 35, FI-40014 Jyv{\"a}skyl{\"a}, Finland}
\newcommand{\karoly}{K\'aroly R\'oberts University College, H-3200 Gy\"ngy\"os, M\'atrai\'ut 36, Hungary}
\newcommand{\kek}{KEK, High Energy Accelerator Research Organization, Tsukuba, Ibaraki 305-0801, Japan}
\newcommand{\korea}{Korea University, Seoul, 136-701, Korea}
\newcommand{\kurchatov}{National Research Center ``Kurchatov Institute", Moscow, 123098 Russia}
\newcommand{\kyoto}{Kyoto University, Kyoto 606-8502, Japan}
\newcommand{\labllr}{Laboratoire Leprince-Ringuet, Ecole Polytechnique, CNRS-IN2P3, Route de Saclay, F-91128, Palaiseau, France}
\newcommand{\lahorelums}{Physics Department, Lahore University of Management Sciences, Lahore 54792, Pakistan}
\newcommand{\lawllnl}{Lawrence Livermore National Laboratory, Livermore, California 94550, USA}
\newcommand{\losalamos}{Los Alamos National Laboratory, Los Alamos, New Mexico 87545, USA}
\newcommand{\lpc}{LPC, Universit{\'e} Blaise Pascal, CNRS-IN2P3, Clermont-Fd, 63177 Aubiere Cedex, France}
\newcommand{\lund}{Department of Physics, Lund University, Box 118, SE-221 00 Lund, Sweden}
\newcommand{\maryland}{University of Maryland, College Park, Maryland 20742, USA}
\newcommand{\mass}{Department of Physics, University of Massachusetts, Amherst, Massachusetts 01003-9337, USA}
\newcommand{\michigan}{Department of Physics, University of Michigan, Ann Arbor, Michigan 48109-1040, USA}
\newcommand{\muenster}{Institut f\"ur Kernphysik, University of Muenster, D-48149 Muenster, Germany}
\newcommand{\muhlenberg}{Muhlenberg College, Allentown, Pennsylvania 18104-5586, USA}
\newcommand{\myongji}{Myongji University, Yongin, Kyonggido 449-728, Korea}
\newcommand{\nagasaki}{Nagasaki Institute of Applied Science, Nagasaki-shi, Nagasaki 851-0193, Japan}
\newcommand{\nara}{Nara Women's University, Kita-uoya Nishi-machi Nara 630-8506, Japan}
\newcommand{\natmephi}{National Research Nuclear University, MEPhI, Moscow Engineering Physics Institute, Moscow, 115409, Russia}
\newcommand{\newmex}{University of New Mexico, Albuquerque, New Mexico 87131, USA}
\newcommand{\nmsu}{New Mexico State University, Las Cruces, New Mexico 88003, USA}
\newcommand{\ohio}{Department of Physics and Astronomy, Ohio University, Athens, Ohio 45701, USA}
\newcommand{\ornl}{Oak Ridge National Laboratory, Oak Ridge, Tennessee 37831, USA}
\newcommand{\orsay}{IPN-Orsay, Univ. Paris-Sud, CNRS/IN2P3, Universit\'e Paris-Saclay, BP1, F-91406, Orsay, France}
\newcommand{\peking}{Peking University, Beijing 100871, P.~R.~China}
\newcommand{\pnpi}{PNPI, Petersburg Nuclear Physics Institute, Gatchina, Leningrad region, 188300, Russia}
\newcommand{\riken}{RIKEN Nishina Center for Accelerator-Based Science, Wako, Saitama 351-0198, Japan}
\newcommand{\rikjrbrc}{RIKEN BNL Research Center, Brookhaven National Laboratory, Upton, New York 11973-5000, USA}
\newcommand{\rikkyo}{Physics Department, Rikkyo University, 3-34-1 Nishi-Ikebukuro, Toshima, Tokyo 171-8501, Japan}
\newcommand{\saispbstu}{Saint Petersburg State Polytechnic University, St.~Petersburg, 195251 Russia}
\newcommand{\saopaulo}{Universidade de S{\~a}o Paulo, Instituto de F\'{\i}sica, Caixa Postal 66318, S{\~a}o Paulo CEP05315-970, Brazil}
\newcommand{\seoulnat}{Department of Physics and Astronomy, Seoul National University, Seoul 151-742, Korea}
\newcommand{\stonybrkc}{Chemistry Department, Stony Brook University, SUNY, Stony Brook, New York 11794-3400, USA}
\newcommand{\stonycrkp}{Department of Physics and Astronomy, Stony Brook University, SUNY, Stony Brook, New York 11794-3800, USA}
\newcommand{\tenn}{University of Tennessee, Knoxville, Tennessee 37996, USA}
\newcommand{\titech}{Department of Physics, Tokyo Institute of Technology, Oh-okayama, Meguro, Tokyo 152-8551, Japan}
\newcommand{\tsukuba}{Center for Integrated Research in Fundamental Science and Engineering, University of Tsukuba, Tsukuba, Ibaraki 305, Japan}
\newcommand{\vandy}{Vanderbilt University, Nashville, Tennessee 37235, USA}
\newcommand{\weizmann}{Weizmann Institute, Rehovot 76100, Israel}
\newcommand{\wigner}{Institute for Particle and Nuclear Physics, Wigner Research Centre for Physics, Hungarian Academy of Sciences (Wigner RCP, RMKI) H-1525 Budapest 114, POBox 49, Budapest, Hungary}
\newcommand{\yonsei}{Yonsei University, IPAP, Seoul 120-749, Korea}
\newcommand{\zagreb}{University of Zagreb, Faculty of Science, Department of Physics, Bijeni\v{c}ka 32, HR-10002 Zagreb, Croatia}
\affiliation{\abilene}
\affiliation{\augie}
\affiliation{\banaras}
\affiliation{\barc}
\affiliation{\baruch}
\affiliation{\bnlcoll}
\affiliation{\bnlphys}
\affiliation{\caucr}
\affiliation{\charlesczech}
\affiliation{\chonbuk}
\affiliation{\ciae}
\affiliation{\cns}
\affiliation{\colorado}
\affiliation{\columbia}
\affiliation{\czechtech}
\affiliation{\dapnia}
\affiliation{\elte}
\affiliation{\ewha}
\affiliation{\fsu}
\affiliation{\gsu}
\affiliation{\hanyang}
\affiliation{\hiroshima}
\affiliation{\howard}
\affiliation{\ihepprot}
\affiliation{\illuiuc}
\affiliation{\inrras}
\affiliation{\instpasczech}
\affiliation{\isu}
\affiliation{\jaea}
\affiliation{\jyvaskyla}
\affiliation{\karoly}
\affiliation{\kek}
\affiliation{\korea}
\affiliation{\kurchatov}
\affiliation{\kyoto}
\affiliation{\labllr}
\affiliation{\lahorelums}
\affiliation{\lawllnl}
\affiliation{\losalamos}
\affiliation{\lpc}
\affiliation{\lund}
\affiliation{\maryland}
\affiliation{\mass}
\affiliation{\michigan}
\affiliation{\muenster}
\affiliation{\muhlenberg}
\affiliation{\myongji}
\affiliation{\nagasaki}
\affiliation{\nara}
\affiliation{\natmephi}
\affiliation{\newmex}
\affiliation{\nmsu}
\affiliation{\ohio}
\affiliation{\ornl}
\affiliation{\orsay}
\affiliation{\peking}
\affiliation{\pnpi}
\affiliation{\riken}
\affiliation{\rikjrbrc}
\affiliation{\rikkyo}
\affiliation{\saispbstu}
\affiliation{\saopaulo}
\affiliation{\seoulnat}
\affiliation{\stonybrkc}
\affiliation{\stonycrkp}
\affiliation{\tenn}
\affiliation{\titech}
\affiliation{\tsukuba}
\affiliation{\vandy}
\affiliation{\weizmann}
\affiliation{\wigner}
\affiliation{\yonsei}
\affiliation{\zagreb}
\author{A.~Adare} \affiliation{\colorado} 
\author{C.~Aidala} \affiliation{\losalamos} \affiliation{\michigan} 
\author{N.N.~Ajitanand} \affiliation{\stonybrkc} 
\author{Y.~Akiba} \affiliation{\riken} \affiliation{\rikjrbrc} 
\author{R.~Akimoto} \affiliation{\cns} 
\author{J.~Alexander} \affiliation{\stonybrkc} 
\author{M.~Alfred} \affiliation{\howard} 
\author{H.~Al-Ta'ani} \affiliation{\nmsu} 
\author{K.R.~Andrews} \affiliation{\abilene} 
\author{A.~Angerami} \affiliation{\columbia} 
% ...
\collaboration{PHENIX Collaboration} \noaffiliation

\date{\today}


\begin{abstract}


The PHENIX experiment at the Relativistic Heavy Ion Collider has 
measured $\phi$ meson production and its nuclear modification in 
asymmetric Cu$+$Au heavy-ion collisions at $\sqrt{s_{NN}}=200$ GeV at both 
forward Cu-going direction ($1.2<y<2.2$) and backward Au-going direction 
($-2.2<y<-1.2$), rapidities.  The measurements are performed via the 
dimuon decay channel and reported as a function of the number of 
participating nucleons, rapidity, and transverse momentum.  In the most 
central events, 0\%--20\% centrality, the $\phi$ meson yield integrated 
over $1<p_T<5$~GeV/$c$ prefers a smaller value, which means a larger 
nuclear modification, in the Cu-going direction compared to the Au-going 
direction.  Additionally, the nuclear-modification factor in Cu$+$Au 
collisions averaged over all centrality is measured to be similar to the 
previous PHENIX result in $d$$+$Au collisions for these rapidities.

\end{abstract}

\pacs{25.75.Dw} 
	
\maketitle

\section{Introduction}

The Relativistic Heavy Ion Collider (RHIC) accelerator and its four 
experiments have previously provided extensive experimental evidence to 
confirm the formation of a deconfined state of nuclear matter, referred to 
as the quark-gluon plasma (QGP), in the initial stages of high-energy 
heavy-ion 
collisions~\cite{Arsene:2004fa,Back:2004je,Adams:2005dq,Adcox:2004mh}. 
Currently, a major objective in the field of high-energy nuclear physics 
is to characterize the properties of the QGP in a quantitative way. The 
$\phi$ meson is a useful probe for studying the QGP properties, because it 
is sensitive to several aspects of the collision, including modifications of strangeness production in bulk matter~\cite{PhysRep-142-167,PhysRevLett-54-1122,Andronic:2014zha}. Due to its small 
inelastic cross section for interaction with nonstrange 
hadrons~\cite{PhysRevLett-54-1122,PhysRevC-49-2198}, the $\phi$ meson is 
less affected by late hadronic rescattering and may reflect the initial 
evolution of the system.  Being composed of a nearly pure strange 
anti-strange ($s\bar{s}$) state, the $\phi$ meson puts additional 
constraints on models of quark recombination in the QGP.

The study of the QGP typically involves comparisons of different 
observables measured in nucleus-nucleus (A$+$B) collisions and in 
proton-proton (\pp) collisions at the same center-of-mass energy. 
Modifications in the A$+$B collisions with respect to \pp collisions could 
be interpreted as being due to the hot nuclear matter (HNM) -- possibly 
QGP -- being produced. However, nuclear modifications could be present in 
the initial state of the collisions even if no QGP is produced. These 
effects, typically referred to as cold nuclear matter (CNM), may include 
the modification of parton distribution functions (PDFs) in a 
nucleus~\cite{Heckman:2007wk}, initial-state energy 
loss~\cite{Vitev:2003xu}, and the Cronin effect, which is often attributed 
to multiple scattering of the incoming parton inside the target 
nucleus~\cite{Cronin:1974zm,Accardi:2003jh}. CNM effects can be probed 
with \dAu collisions. PHENIX has previously measured $\phi$ meson 
production in \dAu collisions at forward, mid- and backward 
rapidities~\cite{Adare:2015vvj}. Suppression was observed in the forward 
($d$-going) direction, where small-x partons from the Au nucleus are 
probed, and an enhancement was seen in the backward (Au-going) direction. 
Similar behavior was previously observed for inclusive charged hadrons and 
open heavy flavor in \dAu 
collisions~\cite{PhysRevLett-94-082302,PhysRevC-90-034903}, potentially 
indicating similar particle production and modification mechanisms.

The rapidity dependence $y$ of particle production in asymmetric 
collisions with a smaller-A projectile and a large-A target, provides a 
way to investigate both hot and cold nuclear-matter effects. 
Previous $J/\psi$ meson 
data in Cu$+$Au collisions~\cite{PhysRevC-90-064908} showed that the ratio 
of forward ($1.2 < y < 2.2$, or Cu-going) to backward ($-2.2 < y < -1.2$, 
or Au-going) $J/\psi$ modification was comparable in both sign and 
magnitude to that expected from CNM effects.  The $\phi$ meson is composed 
of lighter closed flavor ($s\bar{s}$) and its production from 1.0 GeV/$c$ 
to 5.0 GeV/$c$ involves a mix of soft and hard processes and would provide 
a link between heavy flavor and lighter mesons. Comparison of the $\phi$ 
meson production in Cu$+$Au and \dAu systems and to $J/\psi$ production in 
Cu$+$Au collisions may shed light on the mixture of HNM and CNM effects 
on $\phi$-meson production.

The production of $\phi$ mesons has already been measured at PHENIX in 
\pp, \dAu, Cu$+$Cu, and \auau at 
midrapidity~\cite{PhysRevC-72-014903,PhysRevC-83-024909,PhysRevD-83-052004} 
and in \pp and \dAu at forward and backward 
rapidities~\cite{Adare:2014mgt,Adare:2015vvj} over a wide range in \pt. 
Previous measurements from \auau and Cu$+$Cu 
collisions~\cite{PhysRevC-83-024909} in a similar momentum range were 
found to be consistent with HNM effects and exhibited large flow 
anisotropies. The STAR Collaboration has also previously measured $\phi$ 
meson production at midrapidity in Cu$+$Cu and \auau 
collisions~\cite{PhysRevLett-99-112301,PhysLettB-673-183}.
$\phi$ meson production has also been measured by the ALICE Collaboration 
at large rapidity in \pp and $p$$+$Pb collisions~\cite{Adam:2015jca} and 
at midrapidity in \pbpb collisions~\cite{PhysRevC-91-024609}.

In this paper, the production of $\phi$ mesons is determined at forward 
and backward rapidities via dimuons reconstructed in the PHENIX muon 
spectrometers in Cu$+$Au collisions at \sqrtsnn = 200~GeV recorded in 
2012.  The particle multiplicity at these rapidities in heavy-ion 
collisions results in large combinatorial backgrounds and produces a 
challenging environment for $\phi$ meson measurements. Previous 
measurements were thus limited to smaller collision species. A procedure 
for removing the background is detailed and a measurement of the $\phi$ 
meson nuclear modification factor $R_{\rm CuAu}$ in Cu$+$Au collisions at 
forward and backward rapidities is presented versus $y$, \pt, and 
the number of participating nucleons.

\section{Experimental setup}

\begin{figure}[!hbt]
\includegraphics[width=1.0\linewidth]{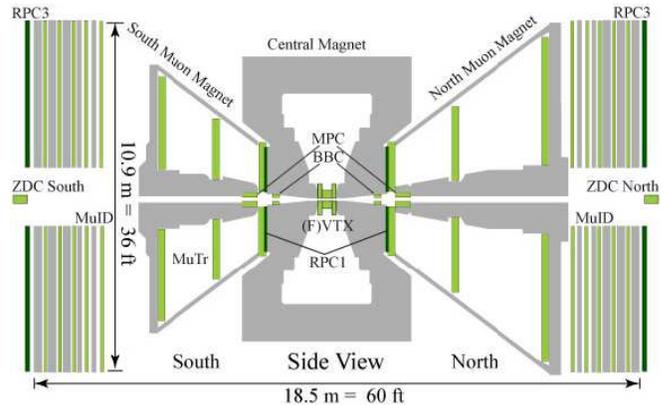}
\caption{\label{fig:PHENIX} (color online) 
The 2012 setup of the PHENIX detector.
}
\end{figure}

The PHENIX detector is described in detail in~\cite{NIMA-499-469}, and a 
schematic of the 2012 setup is shown in Fig.~\ref{fig:PHENIX}. This 
analysis uses the dimuon decay channel of the $\phi$ meson. The detectors 
relevant for this measurement are forward and backward muon 
spectrometers~\cite{NIMA-499-537}, the two beam-beam counters 
(BBCs)~\cite{Allen2003549}, the silicon vertex tracker 
(VTX)~\cite{Taketani:2010zz}, and the forward silicon vertex detector 
(FVTX)~\cite{Aidala:2013vna}.

This study used minimum bias (MB) events triggered by the BBCs.  The BBCs 
comprise two arrays of 64 \v{C}erenkov counters covering the 
pseudorapidity range $3.1<|\eta|<3.9$.  The MB trigger required two or 
more counters firing on each side and a $z$-vertex selection around the 
nominal center of the detector acceptance~\cite{PhysRevC-90-064908}. The 
MB trigger fired on 93$\pm$3\% of the 5.2$\pm$0.2 b total inelastic Cu+Au 
cross section. In this case, the $z$-vertex was measured by the BBCs with 
a resolution of $\sigma_z{\approx}$0.5--2.0~cm, depending on the event 
multiplicity. 

The collision point is determined in $x$, $y$ and $z$ by the two vertex 
detectors, VTX and FVTX, with a resolution of better than 100 microns. The 
VTX and FVTX detectors were installed in 2011 and 2012 to provide precise 
particle vertexing and tracking in the central and forward/backward 
rapidities. Covering approximately the same rapidity range as the existing 
muon spectrometers, the FVTX is composed of two endcaps, each with four 
stations that are perpendicular to the beamline and composed of silicon 
mini-strip sensors that have a 75 micron pitch in the radial direction and 
lengths in the $\phi$ direction varying from 3.4 mm to 11.5 mm. The VTX, 
which surrounds the collision region at PHENIX, comprises four layers of 
silicon sensors. The inner two layers and outer two layers are composed of 
30 pixel ladders and 44 stripixel ladders, respectively.

The muon system is separated into the north and south muon arms.  
Each arm comprises four subcomponents: an absorber material, a magnet, a 
muon tracker (MuTr), and a muon identifier (MuID).  Initially, the 
absorbers were composed of 19 cm copper and 60 cm iron, but an additional 
36.2 cm of stainless steel was added in 2010 to help decrease the hadronic 
background. Following the absorber in each muon arm is the MuTr, which 
comprises three sets of cathode strip chambers in a radial magnetic 
field with an integrated bending power of 0.8 T$\cdot$m.  The final 
component is the MuID, which comprises five alternating steel absorbers 
and Iarocci tubes to further reduce the number of punch-through hadrons 
that can be mistakenly identified as muons. The backplates of the magnets 
provide the first absorber layer for the muon identifier systems. The 
backplate of the south muon magnet is 10 cm shorter than the backplate of 
the north muon magnet, resulting in less total absorber material in the 
south arm than the north arm, and thus a slightly different momentum 
acceptance. The muon spectrometers cover the pseudorapidity range $1.2 < | 
\eta | < 2.2$ over the full azimuth. Muon candidates are identified by 
reconstructed tracks in the MuTr matched to MuID tracks, where at least 
one of the tracks from a pair of muon candidates in the same event 
penetrates through to the last MuID plane. The minimum momentum needed for 
a muon to reach the last MuID plane is $\sim$3 GeV/$c$.

\section{Data Analysis}
\subsection{Dataset and quality cuts}

In this analysis, $\phi$ meson candidates are selected from two 
reconstructed muons in the RHIC Cu$+$Au dataset from 2012. The $\phi$ 
meson invariant yields are then measured and used to calculate the nuclear 
modification factor $R_{\rm CuAu}$, which is compared to results from 
other systems.  For this analysis, 4.73 billion 
($\mathcal{L}=0.97$~nb$^{-1}$) sampled MB events were used within 
$\pm$10~cm $z$-vertex and 0\%--93\% centrality. The total inelastic cross 
section for Cu$+$Au collisions at 200~GeV was estimated by a Glauber 
simulation to be 5.2$\pm$0.2 b.

A set of quality assurance cuts is applied to the data to select good muon 
candidates and improve the signal-to-background ratio. These cuts are 
summarized in Table~\ref{tab:Cuts}. The collision z-vertex is required to 
be within $\pm$10 cm of the center of the interaction region along the 
beam direction, as measured with the BBCs. The MuTr tracks are required to 
match the MuID tracks at the first MuID layer in both position and angle. 
In addition, only dimuon candidates in which at least one track penetrated 
to the final MuID layer are selected.  Furthermore, the track is required 
to have greater than a minimum number of possible hits in the MuTr and 
MuID, and a maximum allowed $\chi^{2}$ is applied to both the track and 
vertex determination. There is a minimum allowed single muon momentum 
along the beam axis, $p_{z}$, which is reconstructed and energy-loss 
corrected at the collision vertex. Finally, this analysis is restricted to 
the dimuon \pt range of $1-5$ GeV/$c$. This limitation is due to the large 
backgrounds and small acceptance at low \pt and small statistics at high 
\pt, preventing signal extraction of the $\phi$ meson. The events are 
sorted into centrality classes using the combined charge from both 
BBCs~\cite{PhysRevC-90-064908}. The number of binary collisions $N_{\rm 
coll}$ and number of participating nucleons $N_{\rm part}$ are extracted 
from a Glauber simulation~\cite{PhysRevC-90-064908}.

\begin{table*}
\caption{
Quality cuts for $\phi$ meson signal extraction in Cu+Au collisions.
}
\begin{ruledtabular} \begin{tabular}{cccl}
Variable & Au-going & Cu-going & Meaning \\ 
\hline
$|z_{\rm vtx}| (cm)$ & $<10$ & $<10$ & 
Collision vertex along the beam direction as measured by the BBCs \\ 
\\
$pDG0$ & $<90$ & $<50$ &
       Track momentum times the spatial difference between \\
\\
(GeV/$c \cdot$ cm)  & & &  the MuTr track and MuID track at the first MuID layer \\ 
\\
$pDDG0$ & $<30$ & $<45$ & 
       Track momentum times the slope difference between \\
(GeV/$c \cdot$ radian) & & &  the MuTr track and MuID track at the first MuID layer \\ 
\\
Track $\chi^2$ & $<5$ & $<10$ & $\chi^{2}/NDF$ of the $\mu$ track \\ 
\\
Lastgap & one track $\geq2$   & one track $\geq2$ &
        Last MuID plane that the $\mu$ track penetrated \\
        & other track $\geq4$   & other track $\geq4$ & \\ 
\\
nidhits & $>$(2$\times$lastgap $-1$) & $>$(2$\times$lastgap $-1$) & 
Number of hits in the MuID, out of the maximum 10 \\ 
ntrhits & $>11$ & $>10$ & 
Number of hits in the MuTr, out of the maximum 16 \\ 
$\chi^{2}_{\rm vtx}$ & $<4$ & $<7$ & 
$\chi^{2}/NDF$ of the dimuon track with the vertex \\
Dimuon \pt (GeV/$c$)& $1-5$ & $1-5$  & 
Transverse momentum of the dimuon pair \\ 
$|p_{z}| (GeV/$c$)$ & $>2.4$ & $>2.5$ & 
Momentum of the $\mu$ along the beam axis \\ 
\label{tab:Cuts}
\end{tabular} \end{ruledtabular}
\end{table*}

\subsection{Background subtraction}

The PHENIX muon spectrometers have a small acceptance for $\phi$ mesons. 
Going from the most peripheral centrality bin, 40\%--93\%, to the most 
central bin, 0\%--20\%, the signal-to-background ratio decreases from 0.28 
to 0.067 in the Cu-going direction ($1.2<y<2.2$) and from 0.37 to 0.090 in 
the Au-going direction ($-2.2<y<-1.2$). Due to the very low 
signal-to-background ratio, particularly in the most central events, the 
background subtraction is of crucial importance. Accordingly, several 
different background subtraction methods were explored and compared.

The invariant mass distribution is formed by combining muon candidate 
tracks of opposite charge. This unlike-sign invariant mass spectrum 
contains the $\phi$, $\rho$ and $\omega$ mesons as well as both 
uncorrelated and correlated backgrounds. The uncorrelated backgrounds come 
from random combinatorial associations of muon candidates, while the 
correlated backgrounds arise from open charm decay (e.g., $D\bar{D}$ where 
both decay semileptonically to muons), open beauty decay, $\eta$ meson and 
$\omega$ meson Dalitz decays and the Drell-Yan process. These correlated 
backgrounds are described in Sec. IIIC.  The uncorrelated combinatorial 
background is accounted for via two methods: (1) like-sign dimuons and (2) 
event mixing.

First, the uncorrelated combinatorial background is estimated through the 
like-sign background subtraction technique, which is generally associated 
with the assumption that the like-sign dimuon pairs come purely from 
combinatorial processes without any correlation between muons. It follows 
that the like-sign distribution can be subtracted from the unlike-sign 
distribution according to the relationship described in Eq.~\ref{eqn:LS1}

\begin{equation}
\label{eqn:LS1}
N_{+-} = FG_{+-} - FG_{\pm\pm},
\end{equation}
where $N_{+-}$ is the uncorrelated background subtracted signal and 
$FG_{+-}$ and $FG_{\pm\pm}$ are the unlike-sign and like-sign dimuon 
pairs, respectively, corresponding to pairs formed within the same event. 
The like-sign distribution $FG_{\pm\pm}$ is normalized to a quantity that 
is more precise and not sensitive to differences in the detector 
acceptance between like-sign and unlike-sign pairs. This background 
normalization is described in Eq.~\ref{eqn:LS2}~\cite{BAGLIN1989471}
\begin{equation}
\label{eqn:LS2}
FG_{\pm\pm} = (FG_{++} + FG_{--}) \frac{2\sqrt{\int FG_{++} \mathrm{d}m \int FG_{--} \mathrm{d}m}}{\int (FG_{++}+FG_{--}) \mathrm{d}m},
\end{equation}
where $m$ is the dimuon invariant mass, and the integration is carried out 
in the range $0.2<m<5.0$ GeV/$c^{2}$.

In parallel to the like-sign technique, the uncorrelated background is 
also estimated through the event mixing technique. In the standard event 
mixing method, muons from different events are randomly associated to 
produce a background distribution of uncorrelated dimuon pairs. Events 
were mixed with partners from within the same 2\%-centrality and 1-cm 
$z$-vertex bins in order to minimize the systematic uncertainties. The 
mixed-event background distributions ($BG$) were generated with about 8 
times higher statistics than the actual background and then normalized to 
match the same-event foreground ($FG$).  The normalization factor also 
accounts for slightly different multiplicities from mixing of slightly 
different events. Although a mass-dependent technique was developed for 
this analysis, a standard event mixing technique is described in advance. 
In previous PHENIX analyses, the normalization factor $\alpha$ was 
calculated as described in Eq.~\ref{eqn:norm1}

\begin{equation}
\label{eqn:norm1}
\alpha = \sqrt{\frac{\int FG_{++} \mathrm{d}m \int FG_{--} 
\mathrm{d}m}{\int BG_{++} \mathrm{d}m \int BG_{--} \mathrm{d}m}},
\end{equation}
where $FG_{++}$ and $FG_{--}$ are the like-sign pairs from the same event 
and $BG_{++}$ and $BG_{--}$ are the like-sign pairs from mixed events.

\begin{figure*}[!hbt]
\includegraphics[width=0.998\linewidth]{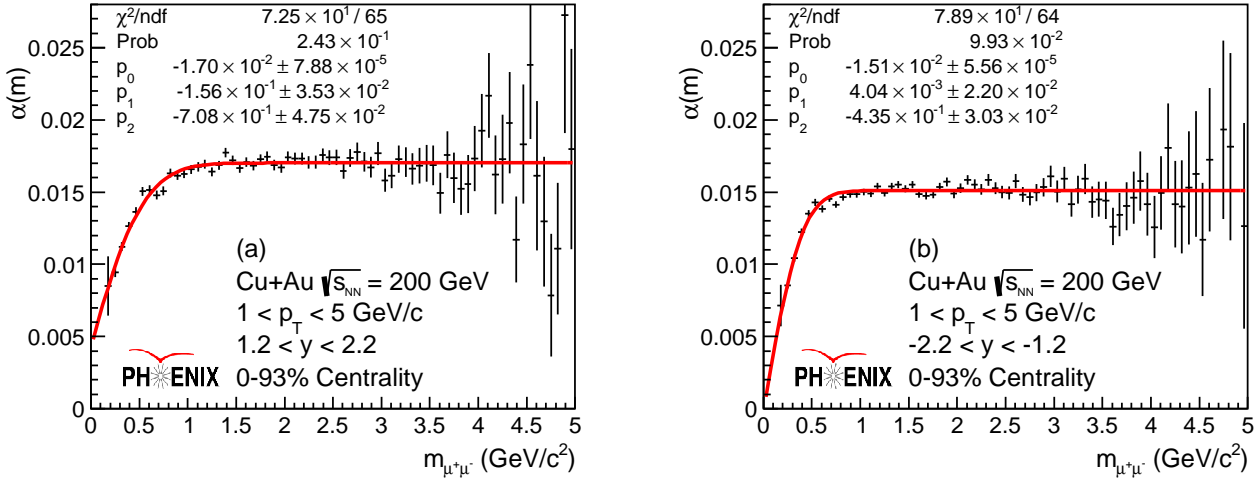}
\caption{\label{fig:alpha} (color online) 
The event mixing normalization factor $\alpha$ versus mass. This factor 
shows a dependence on mass, particularly in the low mass region. The error 
function, which was used to fit $\alpha$, can also be seen in the plot 
along with the fit parameters and goodness of fit.}
\end{figure*}

After subtracting and fitting the resonances as well as the remaining 
correlated background, the yields from mixed-event background subtraction 
are consistent with the yields from the like-sign technique within 
statistical uncertainties. The event mixing technique is used in this 
analysis due to the statistical limitations of the like-sign technique. 
The differences between the like-sign and event mixing techniques are used 
to determine one component of the systematic uncertainty on the yield, as 
described later in Sec. IIIF.

In this method, each term in the square root of Eq.~\ref{eqn:norm1} was 
integrated over all mass, introducing a mass-independent normalization 
factor~\cite{PhysRevC-90-064908,eventmixing}. Dimuons from same events are 
less likely to be reconstructed in close proximity to each other than 
those in mixed events, resulting in a larger relative number of 
mixed-event dimuons at low mass, where the opening angle is small, than at 
higher mass. Therefore, the normalization factor, which is simply a ratio 
of the like-sign same-event dimuons to like-sign mixed-event dimuons, 
drops at lower masses.  Because this normalization factor depends on mass, 
particularly in the $\phi$ meson region, it became necessary to introduce 
a mass-dependent normalization, as described in Eq.~\ref{eqn:norm2}, 
rather than the more commonly used mass-integrated normalization from 
Eq.~\ref{eqn:norm1}.

\begin{equation}
\label{eqn:norm2}
\alpha(m) = \sqrt{\frac{FG_{++}(m)FG_{--}(m)}{BG_{++}(m)BG_{--}(m)}}
\end{equation}

This mass-dependent normalization factor is then fit as a function of 
mass, and the fit function -- rather than the integrated normalization 
factor -- is multiplied to the unlike-sign mixed-event background to get 
the normalized background spectrum $BG^{\rm normalized}_{+-}$,

\begin{equation}
\label{eqn:norm3}
BG^{\rm normalized}_{+-}(m) = \alpha(m) \times BG_{+-}(m).
\end{equation}

Several fitting functions were tested, including a polynomial and an error 
function. The error function, which is used in the final analysis, is 
described in Eq.~\ref{eqn:erf}, where $g$($m$) is the error function and 
$p_{0}$, $p_{1}$ and $p_{2}$ are free parameters of the fit. A plot of the 
normalization factor as a function of mass fit with an error function is 
shown in Fig.~\ref{fig:alpha}.

\begin{equation}
\label{eqn:erf}
g(m) = p_{0} \times \mathrm{Erf}(\frac{m-p_{1}}{p_{2}})
\end{equation}

\begin{figure*}[!hbt]
\includegraphics[width=0.998\linewidth]{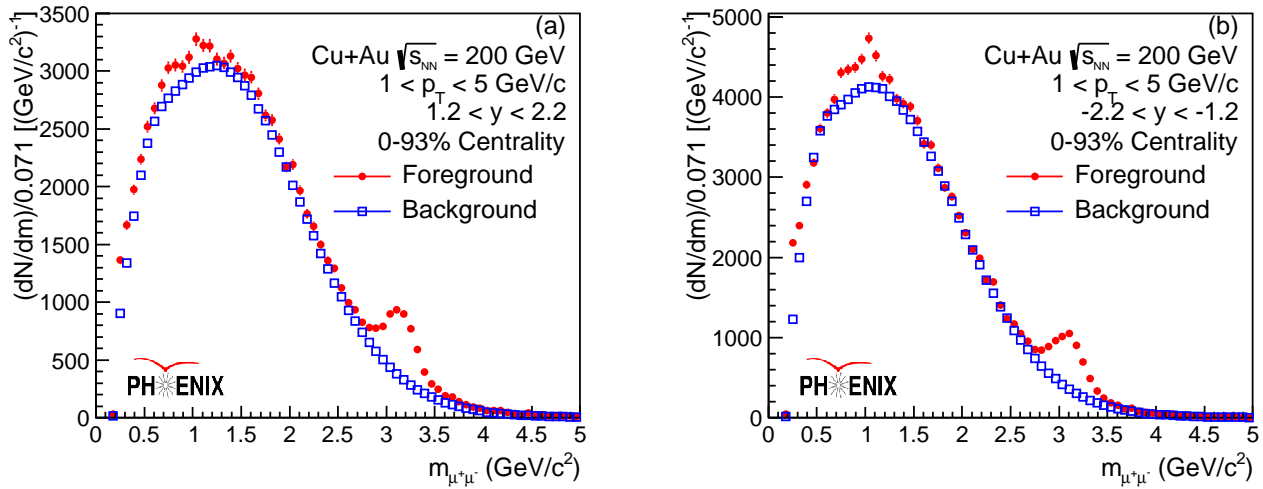}
\caption{\label{fig:mass} (color online) 
The unlike-sign spectra and combinatorial background described with event 
mixing for $1.2 < y < 2.2$ (Cu-going direction) and $-2.2 < y < -1.2$ 
(Au-going direction). The $\rho+\omega$, $\phi$ and $J/\psi$ peaks are 
clearly visible before background subtraction. The mass bin width is 71 
MeV as marked on the vertical axis.}
\end{figure*}

The application of event mixing to describe and subtract backgrounds in 
the $\phi$ meson mass region is shown in Fig.~\ref{fig:mass}, where the 
open squares represent the mixed-event background and the closed circles 
are the unlike-sign spectrum. Before background subtraction, the 
$\rho+\omega$, $\phi$ and $J/\psi$ peaks are clearly seen.

\subsection{Signal extraction and correlated background}

\begin{figure*}[!hbt]
\includegraphics[width=0.998\linewidth]{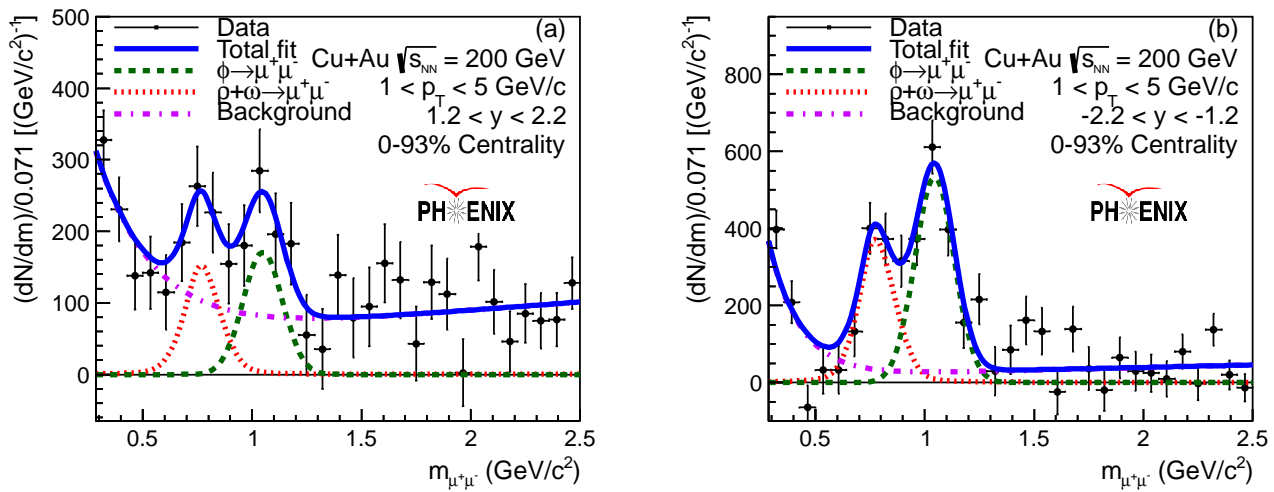}
\caption{\label{fig:mass2} (color online) 
The dimuon mass spectra for $1.2 < y < 2.2$ (Cu-going direction) and $-2.2 
< y < -1.2$ (Au-going direction) after subtracting mixed events and 
fitting the $\phi$ and $\rho+\omega$ peaks and the remaining correlated 
background. The mass bin width is 71 MeV as marked on the vertical axis.
}
\end{figure*}

After the mixed-event background subtraction, there is still some 
correlated background remaining. In previous PHENIX analyses, it was shown 
that heavy flavor (charm and beauty) contributions were negligible in the 
$\phi$ meson mass region for \pp and \dAu collisions at 
200~GeV~\cite{Adare:2014mgt,Adare:2015vvj}. Simulation studies showed that 
$\eta$ meson Dalitz decays are one possible contributor to the correlated 
background. The correlated background is well described by the function in 
Eq.~\ref{eqn:corrfit}

\begin{equation}
\label{eqn:corrfit}
f(m)= \mathrm{exp}(a \cdot m)+b+c \cdot m,
\end{equation}
where $a$, $b$ and $c$ are free parameters of the fit $f$($m$). 
Accordingly, the correlated background in real data are also fit with the 
function described in Eq.~\ref{eqn:corrfit}, as shown in 
Fig.~\ref{fig:mass2}, where the mass distribution after mixed-event 
background subtraction is shown. Several other fit functions and fit 
ranges were tested and used to estimate a systematic uncertainty.

The $\phi$ and $\omega$ meson signals are each described by a Gaussian and 
the signal from the $\rho$ meson by a Breit-Wigner distribution, as shown 
in Fig.~\ref{fig:mass2}, along with the correlated background description.  The $\phi$ meson mass resolution is $\sim$90~MeV/$c^{2}$.  The PHENIX muon arms are not able to resolve the $\rho$ and $\omega$ peaks 
separately, so a combined fit is made. All fit parameters are constrained 
but allowed to vary, except the ratio of the yield of $\rho$ mesons to 
that of $\rho+\omega$, which is set as a constant based on the expected 
ratio between their cross sections and branching ratios.
The data are binned as a function of \pt, $y$ and centrality over the 
range $1 < p_{T} < 5$ GeV/$c$, $1.2 < |y| < 2.2$, and 0\%--93\% centrality.

\subsection{Detector acceptance and reconstruction efficiency}

The product of detector acceptance and reconstruction efficiency, 
$A\varepsilon_{\rm rec}$, of dimuon decays of $\phi$ mesons is determined 
by the full event reconstruction of the $\phi$ meson signal obtained from 
\pythia 6.42~\cite{Phys-commun-135-238}, run through a full 
GEANT3~\cite{geant} simulation of the 2012 PHENIX detector setup, and 
embedded in the MB real-data background. The embedded simulated 
events are then reconstructed in the same manner as data with the same 
cuts applied as in the real data analysis. The background subtraction and 
signal extraction are also handled in the exact same manner as in real 
data. The $A\varepsilon_{\rm rec}$ is then calculated as the number of 
reconstructed $\phi$ meson candidates divided by the number of $\phi$ 
mesons generated in \pythia, both within an appropriate kinematic bin. As 
previously mentioned, the south arm has a smaller amount of absorber 
material, causing a larger acceptance in the south arm (Au-going 
direction) than in the north arm (Cu-going direction). In addition, the 
$A\varepsilon_{\rm rec}$ has a centrality and \pt dependence. 
Specifically, for the lower \pt bin (1-2.5 GeV/$c$), $A\varepsilon_{\rm 
rec} = 1.21 \times 10^{-3}$ in the Cu-going direction and $1.86 \times 
10^{-3}$ in the Au-going direction, while for the higher \pt bin (2.5-5 
GeV/$c$), $A\varepsilon_{\rm rec} = 1.69 \times 10^{-2}$ in the Cu-going 
direction and $1.81 \times 10^{-2}$ in the Au-going direction.  The 
centrality dependence is not as strong, with the values going from 
$A\varepsilon_{\rm rec} = 2.23 \times 10^{-3}$ in the Cu-going direction 
and $2.37 \times 10^{-3}$ in the Au-going direction at 0\%--20\% centrality 
to $A\varepsilon_{\rm rec} = 2.41 \times 10^{-3}$ in the Cu-going 
direction and $3.83 \times 10^{-3}$ in the Au-going direction at 40\%--93\% 
centrality.

\subsection{Invariant yields and nuclear modification factors}

The invariant yield is calculated according to the relation:

\begin{equation}
\label{eqn:bdndydpt}
BR\frac{\mathrm{d}^{2}N}{\mathrm{d}y\mathrm{d}p_{T}}=
\frac{1}{\Delta y\Delta p_{T}}\frac{N}{A\varepsilon_{\rm rec}N_{\rm evt}},
\end{equation}
where $BR$ is the branching ratio to dimuons 
($BR(\phi\rightarrow\mu^{+}\mu^{-}) = 
(2.89\pm0.19)\times10^{-4}$~\cite{PhysRevD-86-010001}), 
$N_{\rm evt}$ is the number of sampled MB events within the relevant 
centrality selection ($N_{\rm evt}=4.73\times10^{9}$ for the 0\%--93\% 
selection), $N$ is the number of observed $\phi$ mesons, and $\Delta y$ 
and $\Delta p_{T}$ are the bin widths in $y$ and \pt, respectively. To 
evaluate the nuclear matter effects on $\phi$ meson production in Cu$+$Au 
collisions, the $\phi$ meson yields in Cu$+$Au collisions are compared to 
those measured in \pp collisions at the same energy after scaling by the 
number of nucleon-nucleon collisions in the Cu$+$Au system, $N_{\rm 
coll}$. This ratio is called the nuclear modification factor $R_{\rm 
CuAu}$, and is defined as:

\begin{equation}
\label{eqn:RAA}
R_{\rm CuAu} = 
\frac{\frac{\mathrm{d}^{2}N_{\rm CuAu}}{\mathrm{d}y\mathrm{d}p_{T}}}{N_{\rm coll}\times\frac{\mathrm{d}^{2}N_{pp}}{\mathrm{d}y\mathrm{d}p_{T}}}.
\end{equation}
The \pp reference data used in the $R_{\rm CuAu}$ are from 
Ref.~\cite{Adare:2014mgt}. Because the rapidity and $p_{T}$ binning in the 
Cu$+$Au analysis differs from that in the \pp analysis, the \pp invariant 
yields were re-measured using the same binning as the Cu$+$Au yields and 
in a manner similar to Ref.~\cite{Adare:2014mgt}. 
The sampled luminosity of the \pp data used in this analysis corresponds 
to $\mathcal{L}=14.1$ pb$^{-1}$~\cite{Adare:2014mgt}.

\subsection{Systematic uncertainties}

\begin{table}
\caption{
Systematic uncertainties included in the invariant yield calculations.
}
\begin{ruledtabular} \begin{tabular}{cccc} 
 Type & Origin                        & Value\\
\hline
  A   & Signal extraction             & 2--31\%\\
\\
  B   & MuID efficiency               & 2\%\\
  B   & MuTr efficiency               & 2\%\\
  B   & $A\varepsilon_{\rm rec}$          & 13\%\\
  B   & $\phi$ candidate selection       & 3\%\\
  B   & Like-sign background subtraction      & 5\%\\
\\
  C  & MB trigger       & 3\%\\
\end{tabular} \end{ruledtabular} 
\label{tab:sysUncer} 
\end{table} 

\begin{table}
\caption{
Systematic uncertainties included in the nuclear modification factor 
calculations.
}
\begin{ruledtabular} \begin{tabular}{cccc} 
 Type & Origin                        & Value\\
\hline
  A   & Signal extraction             & 2--31\%\\
  A   & \pp reference (integrated centrality only)      & 5--13\%\\
\\
  B   & MuID efficiency               & 4\%\\
  B   & MuTr efficiency               & 2\%\\
  B   & $A\varepsilon_{\rm rec}$          & 13\%\\
  B   & $\phi$ candidate selection       & 3\%\\
  B   & Like-sign background subtraction      & 5\%\\
  B   & $N_{\rm coll}$ (centrality bins only)      & 5--10\%\\
\\
  C  & MB trigger       & 10\%\\
  C   & $N_{\rm coll}$ (integrated centrality only)      & 5\%\\
  C   & \pp reference (centrality bins only)      & 11\%\\
\end{tabular} \end{ruledtabular} 
\label{tab:sysUncer2} 
\end{table} 

The systematic uncertainties associated with this measurement are 
categorized as Type-A, Type-B or Type-C. Type-A refers to point-to-point 
uncorrelated uncertainties that allow the data points to move 
independently with respect to one another. They are added in quadrature 
with the statistical uncertainties and represented on the plots as an 
error bar. Type-B uncertainties are correlated point-to-point, which means 
the points move coherently. All sources of Type-B uncertainty are added in 
quadrature and displayed as boxes around the data points. Finally, Type-C 
refers to the global uncertainties which allow the data points to move 
together by an identical multiplicative factor. The Type-C uncertainties 
are given in the legends of the plots.

Several systematic uncertainties are evaluated for this analysis. For the 
signal extraction uncertainty, different fits and parameters are tested 
for the background normalization factor, the correlated background, the 
$\rho+\omega$ signal, and the $\phi$ meson signal. This is done separately 
for each kinematic bin, and a 2-31\% systematic uncertainty is assigned, 
with the largest uncertainty on yields extracted from the most central 
events. This is because the high multiplicity in central collisions results in large combinatorial backgrounds and a very small signal-to-background ratio. It is important to note here that the signal extraction 
uncertainty was primarily dominated by the fluctuations in the correlated 
background. The \pp reference uncertainty comes from the uncertainty on 
the $\phi$ yields in the \pp reference~\cite{Adare:2014mgt}. There is a 
4\% systematic uncertainty from the MuID efficiency and a 2\%
uncertainty from the MuTr efficiency in \pp 
collisions~\cite{Adare:2014mgt}. In Cu$+$Au collisions, the MuTr 
efficiency uncertainty remains the same, while the MuID efficiency 
uncertainty drops down to 2\%~\cite{PhysRevC-90-064908}. For the 
$A\varepsilon_{\rm rec}$ uncertainty, the \pt and $y$ distributions in 
\pythia are changed to match the slope of the distributions in real data, 
and allowed to vary over the range of the error bars in data, yielding a 
13\% systematic uncertainty. Real data and simulation inconsistencies in 
each of the muon identification cuts listed in Table~\ref{tab:Cuts} are 
also evaluated. They can affect the yields by 3\%, which is assigned as 
a systematic uncertainty on the $\phi$ meson candidate selection. The 
like-sign background subtraction uncertainty of 5\% comes from 
differences in the yields when using the like-sign method or the event 
mixing method. The $N_{\rm coll}$ uncertainty of 5--10\% arises from the 
fact that $N_{\rm coll}$ carries a statistical uncertainty itself. 
Finally, the MB trigger efficiency uncertainty was 10\% in the \pp 
reference~\cite{Adare:2014mgt} and 3\% in Cu$+$Au 
collisions~\cite{PhysRevC-90-064908}. All of these systematic 
uncertainties are tabulated in Tables~\ref{tab:sysUncer} 
and~\ref{tab:sysUncer2}.

\section{Results}

\begin{table*}
\caption{Invariant yield as a function of centrality for $1 < p_{T} < 5$ 
GeV/$c$ and $1.2 < |y| < 2.2$. The first value represents the statistical 
and Type-A systematic uncertainties, while the second is the systematic 
uncertainty of Type-B. An additional $\pm$3\% Type-C global systematic 
uncertainty also applies to the yields. The last column summarizes the 
forward/backward ratio shown in Fig.~\ref{fig:N2SNpart}. The 
forward/backward ratio has no Type-C systematic uncertainty.}
\begin{ruledtabular} \begin{tabular}{ccccc} 
 Centrality Bin & $\langle N_{\rm part}\rangle$ & $BR\frac{dN}{dy}$ (Cu-going)
& $BR\frac{dN}{dy}$ (Au-going) & forward/backward ratio\\
\hline
  0\%--20\% & $154.8 \pm 4.1$  & $(7.3 \pm 7.5 \pm 1.1) \times 10^{-5}$            
& $(3.4 \pm 1.0 \pm 0.5) \times 10^{-4}$ & $0.2 ^{+0.3}_{-0.2} \pm <0.1$\\ 
  20\%--40\% & $80.4 \pm 3.3$  & $(1.2 \pm 0.3 \pm 0.2) \times 10^{-4}$             
& $(1.2 \pm 0.3 \pm 0.2) \times 10^{-4}$ & $1.0 ^{+0.4}_{-0.3} \pm 0.1$\\ 
  40\%--93\%  & $19.5 \pm 0.5$ & $(1.5 \pm 0.6 \pm 0.2) \times 10^{-5}$             
& $(2.7 \pm 0.7 \pm 0.4) \times 10^{-5}$ & $0.6 ^{+0.4}_{-0.3} \pm 0.1$\\ 
\end{tabular} \end{ruledtabular}
\label{tab:bdndy1}
\end{table*}

\begin{table*}
\caption{Invariant yield as a function of \pt for 0\%--93\% centrality and 
$1.2 < |y| < 2.2$. The first error represents the statistical and Type-A 
systematic uncertainties, while the second is the systematic uncertainty 
of Type-B. An additional $\pm$5.8\% Type-C global systematic uncertainty 
also applies.}
\begin{ruledtabular} \begin{tabular}{cccc} 
 $p_{T}^{\rm min}$ & $p_{T}^{\rm max}$ & 
$BR\frac{d^{2}N}{dydp_{T}}$ (Cu-going) & 
$BR\frac{d^{2}N}{dydp_{T}}$ (Au-going) \\
(GeV/$c$) & (GeV/$c$) & 
(GeV/$c$)$^{-1}$ & 
(GeV/$c$)$^{-1}$ \\
\hline
  $1.0$ & $2.5$     & $(2.7 \pm 0.8 \pm 0.4) \times 10^{-5}$             & $(5.5 \pm 0.9 \pm 0.8) \times 10^{-5}$\\
  $2.5$ & $5.0$     & $(1.8 \pm 1.0 \pm 0.3) \times 10^{-7}$             & $(4.1 \pm 1.0 \pm 0.6) \times 10^{-7}$\\
\end{tabular} \end{ruledtabular}
\label{tab:bdndy2}
\end{table*}

\begin{table*}
\caption{Invariant yield as a function of rapidity for 0\%--93\% centrality 
and $1 < p_{T} < 5$ GeV/$c$. The first error represents the statistical 
and Type-A systematic uncertainties, while the second is the systematic 
uncertainty of Type-B. An additional $\pm$5.8\% Type-C global systematic 
uncertainty also applies.}
\begin{ruledtabular} \begin{tabular}{cccc} 
 $|y|^{\rm min}$ & $|y|^{\rm max}$ & $BR\frac{dN}{dy}$ (Cu-going)                        & $BR\frac{dN}{dy}$ (Au-going)\\
\hline
1.8 & 2.2   &  $(6.4 \pm 3.1 \pm 0.9) \times 10^{-5}$             & $(1.1 \pm 0.2 \pm 0.2) \times 10^{-4}$\\
  1.2 & 1.8  & $(5.3 \pm 2.3 \pm 0.8) \times 10^{-5}$             & $(1.1 \pm 0.3 \pm 0.2) \times 10^{-4}$\\
\end{tabular} \end{ruledtabular}
\label{tab:bdndy3}

\end{table*}

\begin{table*}
\caption{Nuclear modification factors as a function of centrality for $1 < 
p_{T} < 5$ GeV/$c$ and $1.2 < |y| < 2.2$. The first error represents the 
statistical and Type-A systematic uncertainties, while the second is the 
systematic uncertainty of Type-B. An additional $\pm$15\% Type-C global 
systematic uncertainty also applies.}
\begin{ruledtabular} \begin{tabular}{cccc} 
Centrality Bin & $ \langle N_{\rm coll} \rangle$ & $R_{\rm CuAu}$ (Cu-going) 
& $R_{\rm CuAu}$ (Au-going)\\
\hline
  0\%--20\% & $313.8 \pm 28.4$  & $0.4 \pm 0.4 \pm 0.1$            
& $1.7 \pm 0.5 \pm 0.3$\\
  20\%--40\% & $129.3 \pm 12.4$  & $1.4 \pm 0.4 \pm 0.3$             
& $1.4 \pm 0.3 \pm 0.3$\\
  40\%--93\%  & $21.6 \pm 1.0$ & $1.1 \pm 0.5 \pm 0.2$             
& $1.9 \pm 0.5 \pm 0.3$\\
\end{tabular} \end{ruledtabular}
\label{tab:RAA1}

\end{table*}

\begin{table*}
\caption{Nuclear modification factors as a function of \pt for 0\%--93\% 
centrality and $1.2 < |y| < 2.2$. The first error represents the 
statistical and Type-A systematic uncertainties, while the second is the 
systematic uncertainty of Type-B. An additional $\pm$11\% Type-C global 
systematic uncertainty also applies.}
\begin{ruledtabular} \begin{tabular}{cccc} 
 $p_{T}^{\rm min}$ (GeV/$c$) & $p_{T}^{\rm max}$ (GeV/$c$) & $R_{\rm CuAu}$ (Cu-going)                        & $R_{\rm CuAu}$ (Au-going)\\
\hline
  1.0 & 2.5    & $1.1 \pm 0.4 \pm 0.2$             & $2.3 \pm 0.4 \pm 0.4$\\
  2.5 & 5.0    & $0.6 \pm 0.4 \pm 0.1$             & $1.4 \pm 0.3 \pm 0.2$\\
\end{tabular} \end{ruledtabular}
\label{tab:RAA2}

\end{table*}

\begin{table*}
\caption{
Nuclear modification factors as a function of rapidity for 0\%--93\% 
centrality and $1 < p_{T} < 5$ GeV/$c$. The first error represents the 
statistical and Type-A systematic uncertainties, while the second is the 
systematic uncertainty of Type-B. An additional $\pm$11\% Type-C global 
systematic uncertainty also applies.
}
\begin{ruledtabular} \begin{tabular}{cccc} 
$|y|^{\rm min}$ & $|y|^{\rm max}$ & $R_{\rm CuAu}$ (Cu-going) & $R_{\rm CuAu}$ (Au-going)\\
\hline
  1.8 & 2.2 & $1.2 \pm 0.6 \pm 0.2$             & $2.1 \pm 0.5 \pm 0.3$\\
  1.2 & 1.8   & $0.7 \pm 0.3 \pm 0.1$             & $1.4 \pm 0.4 \pm 0.2$\\
\end{tabular} \end{ruledtabular}
\label{tab:RAA3}

\end{table*}

The invariant yields for $1<p_{T}<5$ GeV/$c$ $\phi$ mesons are calculated 
as a function of centrality, $y$ and \pt as described in 
Eq.~\ref{eqn:bdndydpt}. The results are summarized in 
Tables~\ref{tab:bdndy1} --~\ref{tab:bdndy3}. Similarly, the nuclear 
modification factors are formed from the invariant yields using 
Eq.~\ref{eqn:RAA} and tabulated in Tables~\ref{tab:RAA1} 
--~\ref{tab:RAA3}.

\begin{figure}[!hbt]
\includegraphics[width=1.0\linewidth]{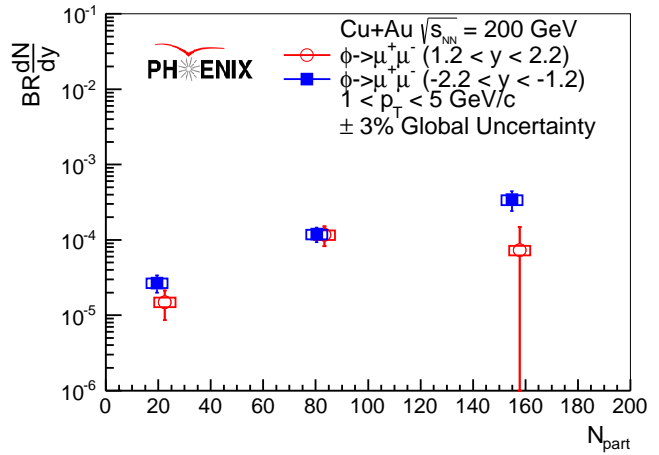}
\caption{\label{fig:NNpart} (color online) 
Invariant yield as a function of the number of participating nucleons for 
$1.2<|y|<2.2$ and $1<p_{T}<5$ GeV/$c$. The centrality bins are 0\%--20\%, 
20\%--40\% and 40\%--93\%, and the data points are placed at the mean 
$N_{\rm part}$ calculated from a Glauber simulation. The data points for 
the Cu-going direction, $1.2<y<2.2$, are shifted along the x-axis to 
$N_{\rm part}$+3 to make the points visible, while the Au-going direction, 
$-2.2<y<-1.2$, remains unshifted. The values are shown in 
Table~\ref{tab:bdndy1}.}
\end{figure}

\begin{figure}[!hbt]
\includegraphics[width=1.0\linewidth]{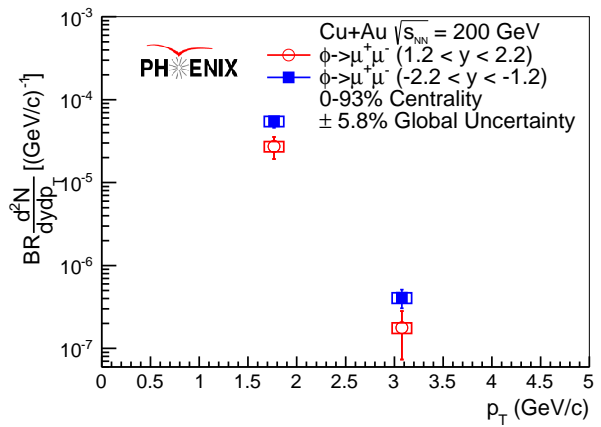}
\caption{\label{fig:NpT} (color online) 
Invariant yield as a function of transverse momentum for $1.2<|y|<2.2$ and 
0\%--93\% centrality. The \pt bins are $1<p_{T}\leq2.5$ and $2.5<p_{T}<5$ 
GeV/$c$, and the data points are placed at the mean \pt of the bin. The 
Cu-going direction corresponds to the forward rapidity, $1.2<y<2.2$, while 
the Au-going direction corresponds to the backward rapidity, 
$-2.2<y<-1.2$. The values are shown in Table~\ref{tab:bdndy2}.}
\end{figure}

\begin{figure}[!hbt]
\includegraphics[width=1.0\linewidth]{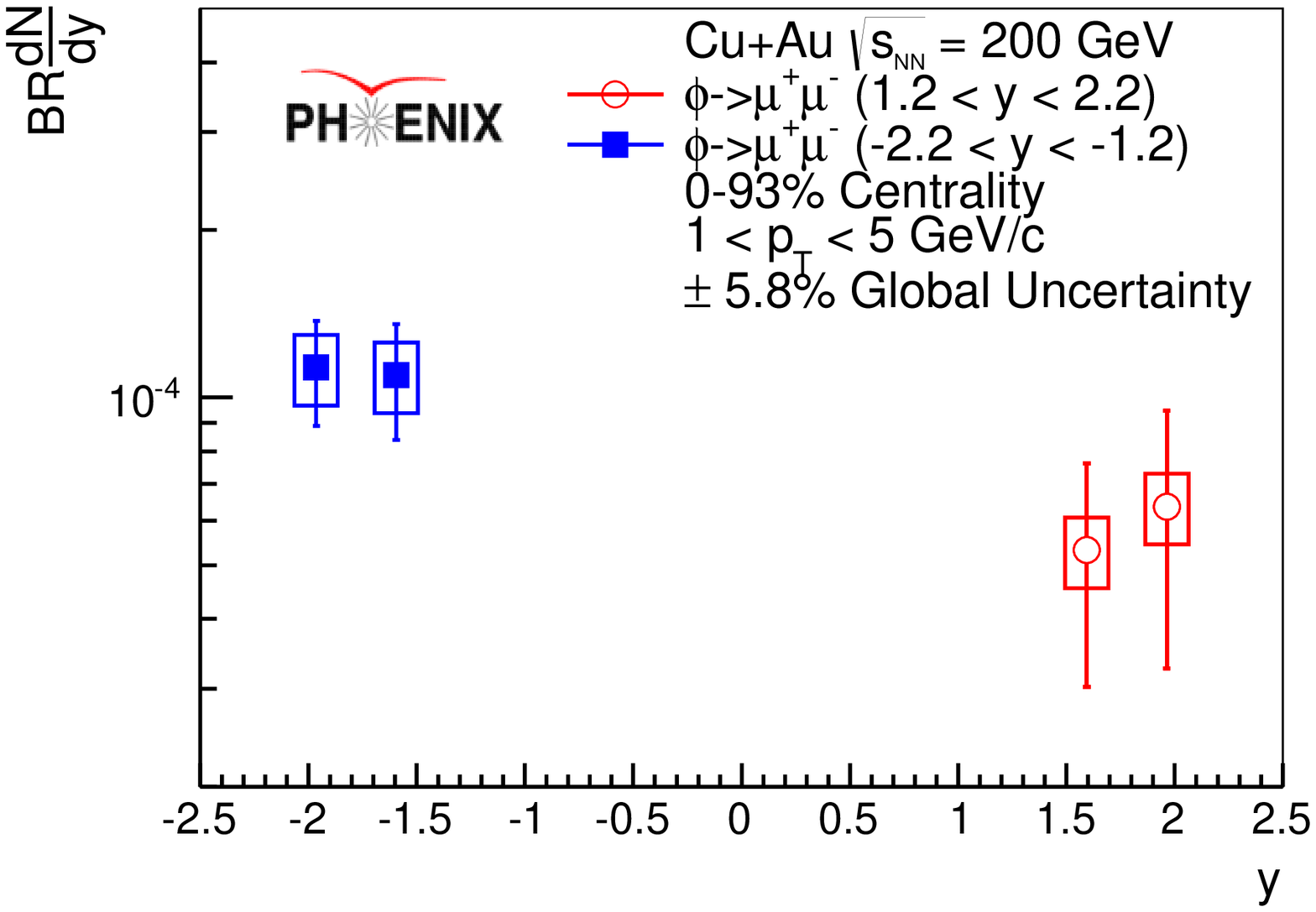}
\caption{\label{fig:Nrap} (color online) 
Invariant yield as a function of rapidity for $1<p_{T}<5$ GeV/$c$ and 
0\%--93\% centrality. The rapidity bins are $1.2<|y|<1.8$ and $1.8<|y|<2.2$ 
and the data points are placed at the mean $y$ of the bin. The Cu-going 
direction covers the region $1.2<y<2.2$, while the Au-going direction 
covers the region $-2.2<y<-1.2$. The values are shown in 
Table~\protect\ref{tab:bdndy3}.}
\end{figure}

Fig.~\ref{fig:NNpart} shows the invariant yield as a function of the 
number of participating nucleons $N_{\rm part}$. In Fig.~\ref{fig:NpT}, 
the dependence of the invariant yield on transverse momentum \pt is shown. 
The invariant yield as a function of rapidity is plotted in 
Fig.~\ref{fig:Nrap}. More $\phi$ mesons are produced in the Au-going 
direction ($-2.2 < y < -1.2$) than in the Cu-going direction($1.2 < y < 
2.2$). This may be explained by the larger multiplicity in the Au-going 
direction coupled with a mixture of both HNM and CNM effects.

Although the invariant yields are interesting on their own, the nuclear 
modification factor is studied in order to evaluate the effects of hot and 
cold nuclear matter on $\phi$ meson production in Cu$+$Au collisions at 
\sqrtsnn = 200~GeV.

\begin{figure}[!hbt]
\includegraphics[width=1.0\linewidth]{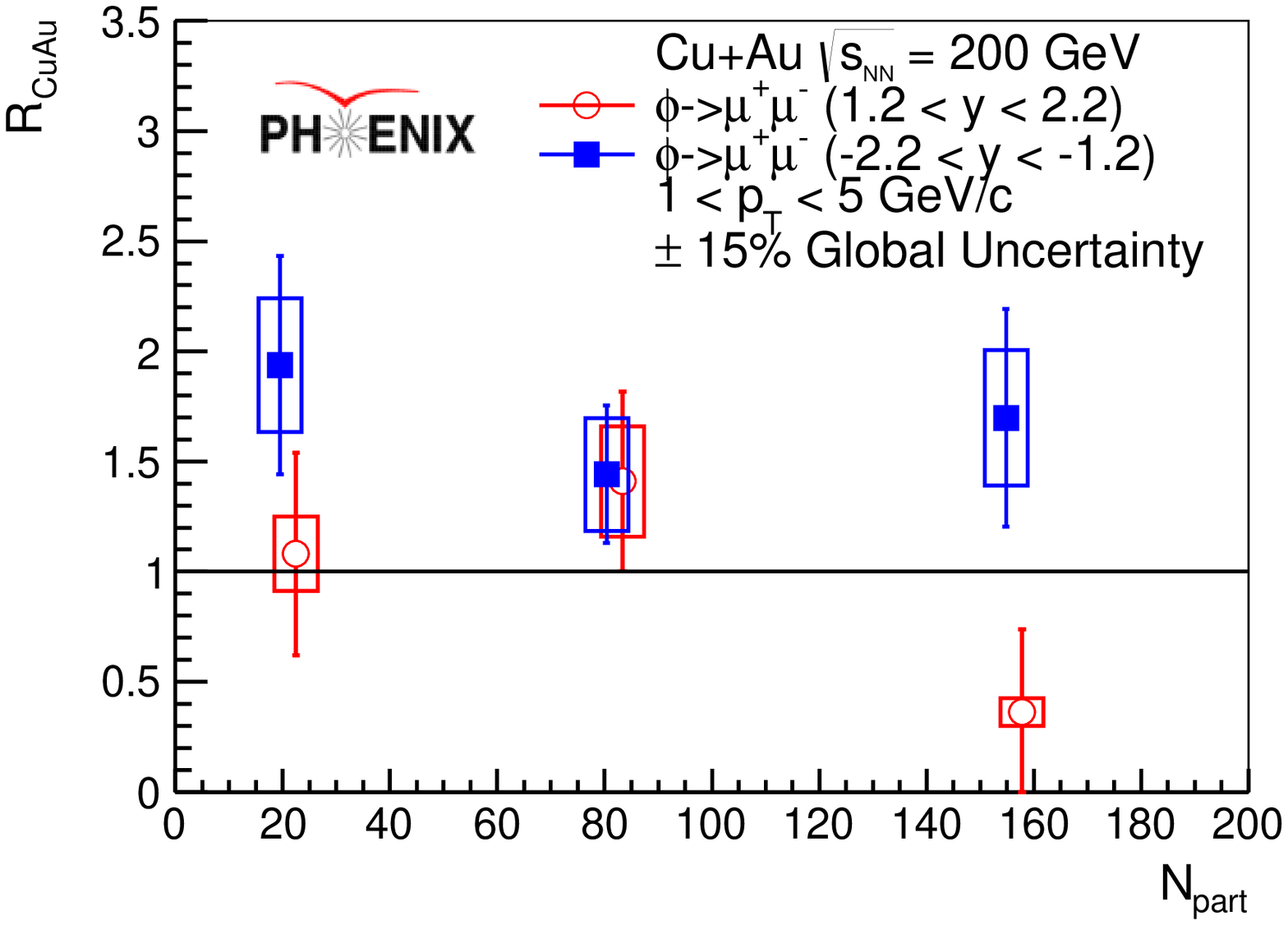}
\caption{\label{fig:RNpart} (color online) 
The nuclear modification factor $R_{\rm CuAu}$ as a function of the number 
of participating nucleons for $1.2<|y|<2.2$ and $1<p_{T}<5$ GeV/$c$. The 
centrality bins are 0\%--20\%, 20\%--40\% and 40\%--93\%, and the data points 
are placed at the mean $N_{\rm part}$ calculated from a Glauber 
simulation. The data points for the Cu-going direction, $1.2<y<2.2$, are 
shifted along the x-axis to $N_{\rm part}$+3 to make the points visible, 
while the data points for the Au-going direction, $-2.2<y<-1.2$, remain 
unshifted. The values are shown in Table~\ref{tab:RAA1}.}
\end{figure}

The nuclear modification factor as a function of $N_{\rm part}$ is shown 
in Fig.~\ref{fig:RNpart}. There is a dependence of $R_{\rm CuAu}$ on both 
centrality and rapidity. In the Au-going direction, the $R_{\rm CuAu}$ is 
greater than unity for all centralities. The rapidity dependence is 
similar to the trend observed by PHENIX for 
$\phi\rightarrow\mu^{+}\mu^{-}$ in \dAu collisions~\cite{Adare:2015vvj} as 
well as measurements made by the ALICE Collaboration at large rapidity in 
$p$$+$Pb collisions at 5.02 TeV at the Large Hadron Collider~\cite{Adam:2015jca}, 
where an enhancement was observed in the Pb-going direction while the 
$p$-going direction was either suppressed or consistent with unity 
depending on the \pt range.

\begin{figure}[!hbt]
\includegraphics[width=1.0\linewidth]{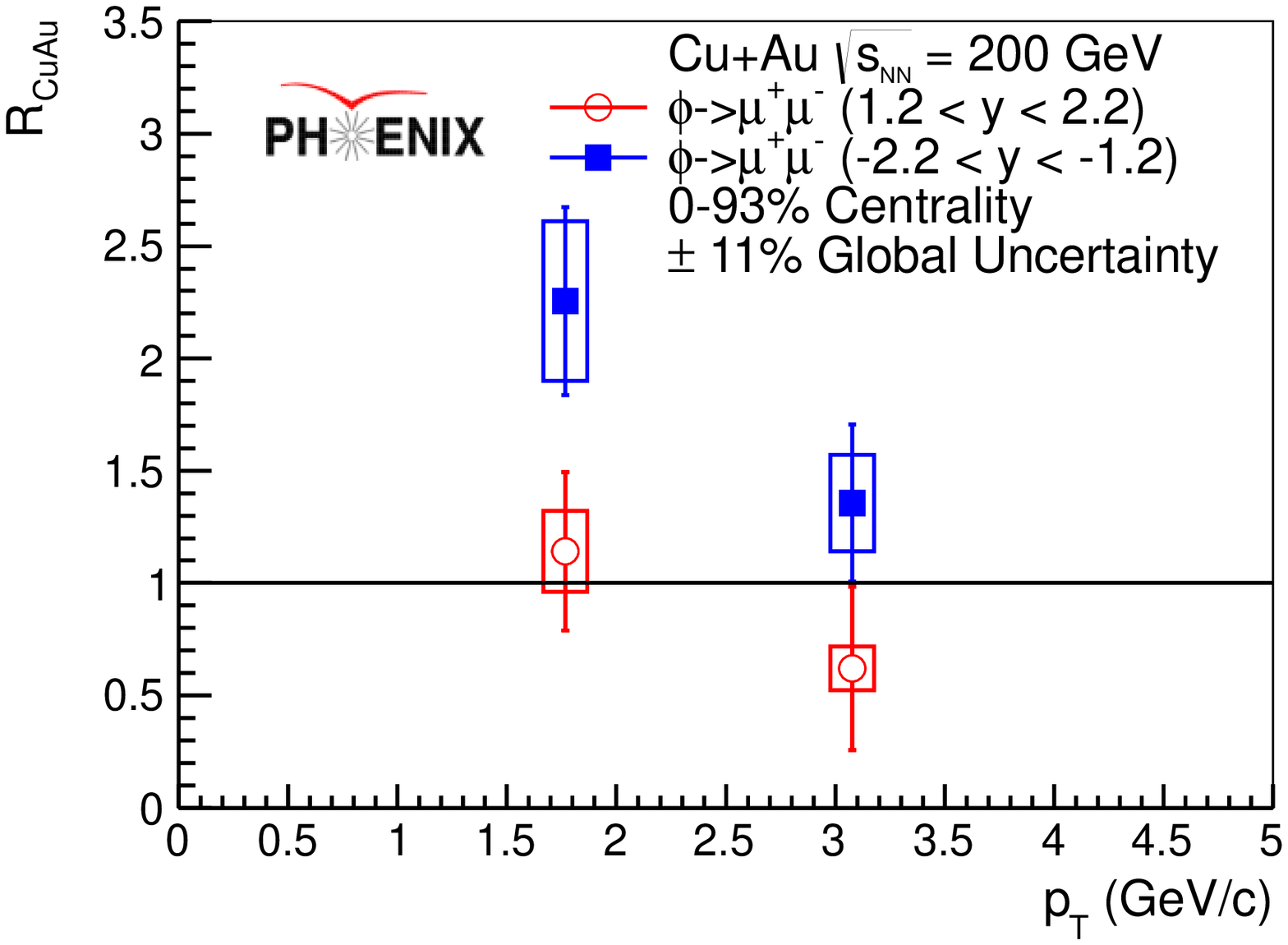}
\caption{\label{fig:RpT} (color online) 
The nuclear modification factor $R_{\rm CuAu}$ as a function of transverse 
momentum for $1.2<|y|<2.2$ and 0\%--93\% centrality. The \pt bins are 
$1<p_{T}\leq2.5$ and $2.5<p_{T}<5$ GeV/$c$, and the data points are placed 
at the mean \pt of the bin. The Cu-going direction corresponds to the 
forward rapidity, $1.2<y<2.2$, while the Au-going direction corresponds to 
the backward rapidity, $-2.2<y<-1.2$. The values are shown in 
Table~\ref{tab:RAA2}.}
\end{figure}

To further understand the relative roles of different nuclear matter 
effects in this collision system, the transverse momentum dependence of 
the nuclear modification factor is shown in Fig.~\ref{fig:RpT}. The data 
points are placed at the mean \pt of the bin. Here, the nuclear 
modification is calculated over integrated centrality, but it should be 
noted that the data are dominated by central collisions. There is an 
enhancement at low \pt in the Au-going direction.  In the Cu-going 
direction, $R_{\rm CuAu}$ is consistent with unity. The enhancement in the 
Au-going direction is similar in scale to that observed in the Au-going 
direction in \dAu collisions~\cite{Adare:2015vvj}, indicating similar 
nuclear modification between the two collision systems.

\begin{figure}[!hbt]
\includegraphics[width=1.0\linewidth]{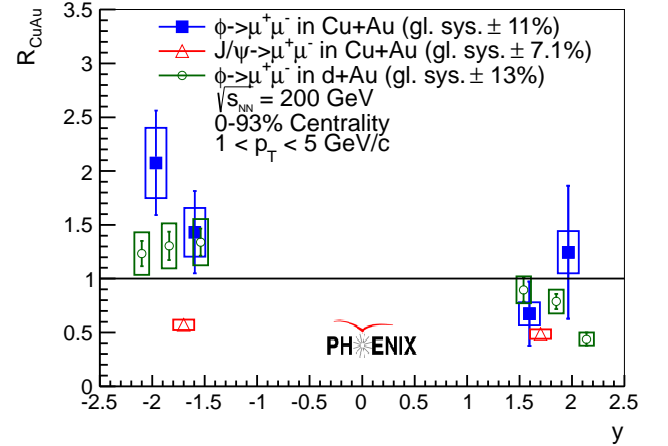}
\caption{\label{fig:Rrap} (color online) 
The nuclear modification factor $R_{\rm CuAu}$ as a function of rapidity 
for $1<p_{T}<5$ GeV/$c$ and 0\%--93\% centrality. The rapidity bins are 
$1.2<|y|<1.8$ and $1.8<|y|<2.2$ and the data points are placed at the mean 
$y$ of the bin. The values are shown in Table~\ref{tab:RAA3}. 
Also included are previous PHENIX results for $\phi$ mesons in \dAu 
collisions~\cite{Adare:2015vvj} represented by open circles and $J/\psi$ 
mesons in Cu$+$Au collisions~\cite{PhysRevC-90-064908} represented by open 
triangles. Positive rapidity, $1.2<y<2.2$, corresponds to the Cu-going and 
d-going directions, while negative rapidity, $-2.2<y<-1.2$, is the 
Au-going direction.}
\end{figure}

\begin{figure}[!hbt]
\includegraphics[width=1.0\linewidth]{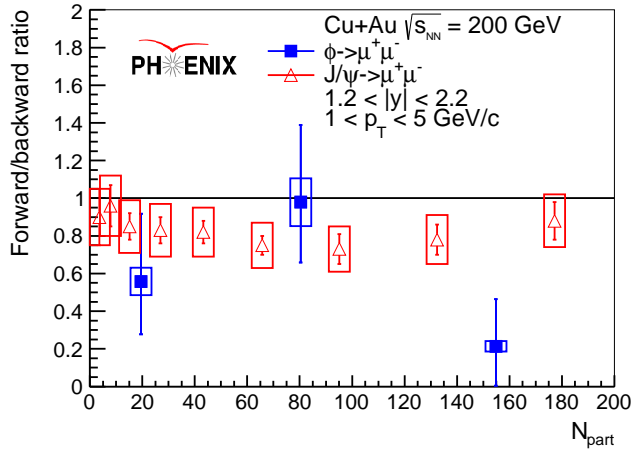}
\caption{\label{fig:N2SNpart} (color online) 
The forward/backward ratio as a function of the number of participating 
nucleons for $1<p_{T}<5$ GeV/$c$ and $1.2<|y|<2.2$. The values are shown 
in Table~\ref{tab:bdndy1}. The Cu-going direction covers positive rapidity, 
$1.2<y<2.2$, while the Au-going direction covers negative rapidity, 
$-2.2<y<-1.2$.}
\end{figure}

Fig.~\ref{fig:Rrap} shows the nuclear modification factor $R_{\rm CuAu}$ 
as a function of $y$ for two rapidity regions, $1.2<|y|<1.8$ and 
$1.8<|y|<2.2$. The data points are placed at the mean $y$ of the bin. As 
in Fig.~\ref{fig:RpT}, the nuclear modification factor is inclusive of 
centrality. The rapidity-dependence of $R_{\rm CuAu}$ is similar to the 
trend observed in previous $\phi$ meson measurements in \pdau collisions. 
In particular, $\phi$ meson production is enhanced in the Au-going 
direction. None of the Cu-going points show significant suppression given 
the statistical uncertainties. For comparison, the PHENIX $J/\psi$ meson 
results in the same Cu$+$Au dataset from Ref.~\cite{PhysRevC-90-064908} 
are also shown in Fig.~\ref{fig:Rrap}. While the closed charm shows 
suppression at both forward and backward rapidity for $1.2<|y|<2.2$, the 
closed strangeness is enhanced at backward rapidity. In Cu$+$Au 
collisions, the $J/\psi$ meson yield is strongly suppressed in the 
Au-going direction compared to the $\phi$ meson yield at the same 
rapidity. This is similar to the differences previously observed between 
$J/\psi$ and $\phi$ meson nuclear modification in \dAu collisions 
\cite{Adare:2015vvj}. These differences could be attributed to a larger 
$J/\psi$ break up cross section, effects in the higher-energy-density 
backward-rapidity region, or changes between soft and hard production 
mechanisms between the two mesons.

The forward and backward differences can be quantified by the ratio of the 
yield values for the forward rapidity (Cu-going direction) to the backward 
rapidity (Au-going direction). Fig.~\ref{fig:N2SNpart} shows the 
forward/backward ratio as a function of participating nucleons for 
$1.2<|y|<2.2$ and $1<p_{T}<5$ GeV/$c$. The Type-C and Type-B systematic 
uncertainties, except for the $A\varepsilon_{\rm rec}$ uncertainty, cancel 
when taking this ratio. The remaining systematic uncertainties are the 
Type-A signal extraction uncertainty and the Type-B $A\varepsilon_{\rm rec}$ 
uncertainty. The difference in suppression between the forward and 
backward rapidity is more noticeable in the most central collisions, 
0\%--20\%. In this centrality bin, the probability of observing the 
forward/backward ratio greater than or equal to unity was found to be 
$p$-value=1.2\%, corresponding to a statistical significance of 
$2.3\sigma$. The particle multiplicity for central collisions should be 
about 20\% higher in the Au-going direction than in the Cu-going 
direction~\cite{PhysRevC-73-14906}, however, the much smaller ratio 
observed may indicate that increased recombination effects or additional 
thermal strangeness production may also occur at higher energy density. In 
central collisions, the forward/backward ratio in $\phi$ production 
($\sim$0.2) is smaller than that in $J/\psi$ production ($\sim$0.8) in 
Cu$+$Au collisions~\cite{PhysRevC-90-064908}.

\section{Summary}

In summary, $\phi$ meson production and its nuclear modification have been 
measured in Cu$+$Au collisions at \sqrtsnn = 200~GeV for $1.2<|y|<2.2$ and 
$1.0<p_{T}<5.0$ GeV/$c$ via the dimuon decay channel. This first 
measurement of $\phi$ meson production and its nuclear modification in a 
heavy-ion system at forward/backward rapidity at RHIC extends measurements 
of $\phi$ from smaller systems, \pp and \dAu, in the forward and backward 
rapidity. The invariant yields and nuclear modification factors have been 
presented here as a function of $N_{\rm part}$, \pt and rapidity.

The $\phi$ meson yields in Cu$+$Au collisions are found to be generally 
smaller in the Cu-going direction than in the Au-going direction. This is 
most pronounced in the most central events, 0\%--20\%, and at low 
momentum, 1.0--2.5~GeV/$c$. In central collisions (0\%--20\%), the 
forward/backward ratio is below unity at a confidence level of~$99\%$. It 
has been shown that these results follow a trend similar to what was seen 
previously at PHENIX in \dAu at the same rapidity and 
energy~\cite{Adare:2015vvj} as well as the ALICE measurement in $p$$+$Pb 
collisions at larger rapidity ($-4.46<y<-2.96$ and $2.03<y<3.53$) and 
higher energy (\sqrtsnn = 5.02~TeV)~\cite{Adam:2015jca}. While this 
agreement could imply a role for CNM effects on $\phi$ production in 
Cu$+$Au collisions, the production of $\phi$ in heavy-ion collisions for 
these kinematics is expected to have substantial contributions from HNM 
effects as which were demonstrated to dominate previous measurements at 
midrapidity for both Cu$+$Cu and \auau 
collisions~\cite{PhysRevC-83-024909}. A competition between CNM and HNM 
production mechanisms appears relevant for $\phi$ production at forward 
rapidity for heavy-ion collisions and a comprehensive description is 
needed from soft and hard physics models.  Although the $\phi$ meson is 
sensitive to both CNM and HNM effects, this study was statistically limited, a factor that also affects the precise determination of the systematic uncertainties. A  high statistics measurement and theory calculations are both needed in order to make conclusions about the various physics processes that might be at play here, including modifications of  strangeness production in bulk matter and quark recombination.


\section*{ACKNOWLEDGMENTS}   

We thank the staff of the Collider-Accelerator and Physics Departments at 
Brookhaven National Laboratory and the staff of the other PHENIX 
participating institutions for their vital contributions.  
We acknowledge support from two Offices within the Office of
Science of the Department of Energy: The Office of Nuclear Physics and 
the Office of Science Graduate Student Research award program, which is 
administered by the Oak Ridge Institute for Science and Education, 
and support from the National Science Foundation,
Abilene Christian University Research Council,
Research Foundation of SUNY, and
Dean of the College of Arts and Sciences, Vanderbilt University
(U.S.A),
Ministry of Education, Culture, Sports, Science, and Technology
and the Japan Society for the Promotion of Science (Japan),
Conselho Nacional de Desenvolvimento Cient\'{\i}fico e
Tecnol{\'o}gico and Funda\c c{\~a}o de Amparo {\`a} Pesquisa do
Estado de S{\~a}o Paulo (Brazil),
Natural Science Foundation of China (P.~R.~China),
Croatian Science Foundation and
Ministry of Science, Education, and Sports (Croatia),
Ministry of Education, Youth and Sports (Czech Republic),
Centre National de la Recherche Scientifique, Commissariat
{\`a} l'{\'E}nergie Atomique, and Institut National de Physique
Nucl{\'e}aire et de Physique des Particules (France),
Bundesministerium f\"ur Bildung und Forschung, Deutscher
Akademischer Austausch Dienst, and Alexander von Humboldt Stiftung 
(Germany),
National Science Fund, OTKA, K\'aroly R\'obert University College,
and the Ch. Simonyi Fund (Hungary),
Department of Atomic Energy and Department of Science and Technology 
(India),
Israel Science Foundation (Israel),
Basic Science Research Program through NRF of the Ministry of Education 
(Korea),
Physics Department, Lahore University of Management Sciences (Pakistan),
Ministry of Education and Science, Russian Academy of Sciences,
Federal Agency of Atomic Energy (Russia),
VR and Wallenberg Foundation (Sweden),
the U.S. Civilian Research and Development Foundation for the
Independent States of the Former Soviet Union,
the Hungarian American Enterprise Scholarship Fund,
and the US-Israel Binational Science Foundation.



%
 
\end{document}